\documentclass[aps, prl, reprint, superscriptaddress]{revtex4-1}
\usepackage{graphicx}
\usepackage{dcolumn}
\usepackage{amsmath}
\newcommand{\etal}{\textit{et al.}}
\newcommand{\fm}{5~\textsuperscript{2}F}
\newcommand{\f}{\fm}
\newcommand{\Kf}{K(\fm)}
\newcommand{\gm}{5~\textsuperscript{2}G}
\newcommand{\g}{\gm}
\newcommand{\Kg}{K(\gm)}
\newcommand{\p}{7~\textsuperscript{2}P}
\newcommand{\sm}{7~\textsuperscript{2}S}

\newcommand{\Ks}{K(\sm)}

\newcommand{\figWidth}{8.6cm}

\begin{document}

\title{Threshold photodetachment in a repulsive potential}

\author{A. O. Lindahl}
\email{anton.lindahl@physics.gu.se}
\affiliation{Department of Physics, University of Gothenburg, 412 96 Gothenburg, Sweden}

\author{J. Rohl\'en}
\affiliation{Department of Physics, University of Gothenburg, 412 96 Gothenburg, Sweden}

\author{H. Hultgren}
\affiliation{Department of Physics, University of Gothenburg, 412 96 Gothenburg, Sweden}
\affiliation{Albert-Ludwigs-Universit\"at, D-79104 Freiburg, Germany}
\author{I. Yu. Kiyan}
\affiliation{Albert-Ludwigs-Universit\"at, D-79104 Freiburg, Germany}

\author{D. J. Pegg}
\affiliation{Department of Physics, University of Tennessee, Knoxville, Tennessee 37996, USA}

\author{C. W. Walter}
\affiliation{Department of Physics and Astronomy, Denison University, Granville, Ohio 43023, USA}

\author{D. Hanstorp}
\email{dag.hanstorp@physics.gu.se}
\affiliation{Department of Physics, University of Gothenburg, 412 96 Gothenburg, Sweden}

\date{\today}

\begin{abstract}
We report on the first experimental observation of a new threshold behavior observed in the \g\ partial channel in photodetachment of K$^-$.  It arises from the repulsive polarization interaction between the detached electron and the residual \Kg\ atom, which has a large negative dipole polarizability.   In order to account for the observation in the \Kg\ channel, we have developed a semiclassical model that predicts an exponential energy dependence for the cross section.  The measurements were made with collinear laser-ion beams and a resonance ionization detection scheme.
\end{abstract}


\maketitle

Inelastic scattering processes involving resonant excitation have provided a wealth of information on bound and quasibound states, whereas less emphasis has been placed on nonresonant or continuum processes. Studies of threshold behavior in such processes are of fundamental interest since they allow us to better understand the nature of the interactions between particles created by fragmentation. For example, Einstein's interpretation \cite{einstein:132} of the frequency dependence of the threshold position in the photoelectric effect  was instrumental in the development of quantum mechanics.
Threshold effects have been shown to be important in diverse areas of physics including atomic and molecular, subatomic and solid state physics \cite{sadeghpour:R93}. Specific examples include nuclear halo states \cite{jonson:1}, evaporative cooling of fermionic atoms \cite{demarco:4208}, molecular dissociation \cite{zhang:203003} and the neutron capture process \cite{rupak:222501}.

In the field of atomic physics, the interpretation of particle scattering and photo-fragmentation experiments is aided by the fact that the interparticle interactions are well understood. One particularly interesting photo-fragmentation process is photodetachment. Here an electron is ejected from a negative ion following the absorption of a photon.
Studies of threshold behaviors in one-electron photodetachment are of fundamental interest. Just above threshold, the detached electron moves very slowly relative to the residual atom and its motion and that of the remaining atomic electrons become highly correlated. In this system, subtle correlation interactions are not masked by far stronger Coulomb forces, which is the case in photoionization.
Wigner showed that the centrifugal potential determines the behavior of the cross section in the threshold region for two-particle breakup reactions  for which the interaction approaches zero faster than $r^{-2}$ \cite{wigner:1002}. The cross section is then given by
$
\sigma_{\mathrm{W}}\sim\,(E_{h\nu}-E_\mathrm{th})^{{\ell+1/2}} = {E_e}^{{\ell+1/2}},
$
where $E_{h\nu}$ is the photon energy, $E_\mathrm{th}$ is the threshold energy and $E_e$ and $\ell$ are the kinetic energy and angular momentum of the detached electron, respectively.

The Wigner law is strictly valid only at threshold. However, good agreement between the Wigner law and measured cross sections are typically observed up to energies on the order of 10~meV above threshold, see for example Refs.~\cite{bilodeau:012505, hotop:762, hotop:2379}.
In one particular case, however, when detachment of an inner-shell electron was investigated, the data followed the Wigner law up to 2~eV above threshold \cite{bilodeau:083001}.
The lowest order correction terms to the Wigner law include the effects of the finite size of the initial state wave function \cite{farley:6286} and the polarizability of the residual atom \cite{omalley:1668}.
Several theoretical  studies of threshold behavior in the presence of an attractive polarization interaction have been made \cite{watanabe:158, omalley:1668}. 
The approximations made in the model by O'Malley \cite{omalley:1668}, however, severely reduce its applicability when the polarizability is large \cite{sandstrom:052707}.
In photodetachment experiments on Li$^-$ and K$^-$, in which the atoms were left in highly polarizable states, a more advanced modified effective range model \cite{watanabe:158} had to be applied to describe the observed threshold behaviors \cite{sandstrom:052707}.

In this Letter we present the results of an experiment on one-electron photodetachment from the K$^-$ ion. Specifically, the partial cross sections for the K(\fm)$ + e^-(\epsilon d)$ and K(\gm)$ + e^-(\epsilon f)$ final state channels have been measured.
The dipole polarizabilities of the K atom in the \f\ and the \g\ states have been calculated to be $3\,936\,137$ and $-3\,097\,696$~a.u.,
respectively  \cite{liu:052715}.
The polarizability of the $4\,^2\!S$ ground state is 307 a.u.
The large polarizabilities arise since the \f\ and \g\} states are energetically close (separated by only 1.3 meV \cite{sansonetti:7}) and thus interact strongly in an external field, while other states are far away.
Because of the large polarizabilities, the polarization potential is expected to play a significant role in determining the behavior of the cross section above both thresholds.
Our measurements exhibit very different behaviors above the two thresholds: a steplike onset in the \f\ channel, while the increase of the cross section in the \g\ channel is very slow. This difference is attributed to the sign difference of the polarizability and thus whether the final state  interaction is attractive or repulsive.
This is the first reported observation of the effect of a strong, repulsive polarization interaction on a photodetachment threshold.

The effects of the attractive and repulsive potentials on the onset of production of electrons in photodetachment are reminiscent of the effects seen in the nuclear $\beta$ decay distributions \cite{blatt:weisskopf}. In $\beta^-$ decay there is a steplike onset at the low-energy end of the electron distribution, in sharp contrast to the suppressed production of low-energy $\beta^+$ particles. This difference is explained by the attractive and repulsive Coulomb interactions, respectively, in the final states.

In order to measure the small partial cross sections for the \Kf\ and \Kg\ channels, the experiment was performed in a collinear laser-ion beam geometry to maximize the sensitivity.
A resonance ionization scheme was used to distinguish the two channels.
The experiment was performed at Gothenburg University Negative Ion Laser Laboratory.
Negative ions of potassium were created from K$_2$CO$_3$ in a sputter ion source and were accelerated to 6 keV energy. The ion beam was mass selected in a magnetic sector before it was bent by an electrostatic quadrupole deflector into an interaction region defined by two 3~mm apertures placed 61~cm apart. This part of the setup has been described in detail by Diehl \etal\ \cite{diehl:053302}. The interaction region is shielded against stray electric fields by means of a stainless steel tube centered along the path of the ions. A SIMION \cite{dahl:3} simulation showed that the residual electric field in the interaction region is substantially smaller than 1 V/cm. An electric field of this strength would result in a mixing of the \f\ and \g\ states that is less than $10^{-4}$, and this effect can thus be neglected. In the interaction region, the ion beam was merged with two copropagating laser beams. A pulsed UV laser was used to photodetach the K$^-$ ions. A pulsed IR laser resonantly excited the residual atoms to specific Rydberg states.
After the interaction region, an inhomogeneous electric field and a position sensitive detector (PSD) served as a Rydberg state analyzer. The analyzer will be described in more detail in a forthcoming publication. Rydberg atoms in different states were field ionized at different positions in the field
so that the resulting positive ions were deflected at different angles into the detector. Information on where the ions hit the PSD was used to distinguish between different Rydberg states and to separate the field-ionized Rydberg atoms from positive ions created in sequential photodetachment-photoionization by two UV photons. Another advantage of the PSD was that it enabled subtraction of diffuse background processes. The electrical field in the Rydberg analyzer also deflected the negative ion beam into a Faraday cup.  For each shot of the two lasers, the data from the PSD were recorded together with the measured wavelength and pulse energy of the UV radiation and the ion current.

The ion current in the interaction region was on the order of 1~nA. Both lasers were Nd:YAG-pumped Optical parametric oscillators delivering about 6~ns pulses with repetition rates of 10~Hz. The specified bandwidths were 0.2~cm$^{-1}$ $\approx$ 25~$\mu$eV. The pulse energy of the IR laser after the chamber was approximately 0.2~mJ, which was sufficient to saturate the resonant transitions to the Rydberg states. The UV energy was approximately 0.7~mJ, far from saturating the nonresonant photodetachment process. In order to investigate the threshold regions, the UV radiation was tuned from  4.290 to 4.352~eV in the ion rest frame. The IR laser was tuned to one of the resonant transitions $\sm \rightarrow 25\,^2\!P$, $\fm \rightarrow 23\,^2\!D$ or
$\gm \rightarrow 22\,^2\!F$, which, in the rest frame, correspond to 562.050, 519.566, and 516.461~meV, respectively \cite{sansonetti:7}.  The  UV pulse arrived at the chamber approximately 50 ns before the IR pulse.

Counts in a selected area of the detector were summed, and the background around the selected area was subtracted. This number was normalized with respect to the ion current and the pulse energy of the UV laser and finally sorted and binned based on the measured UV wavelength.
The measured photon energies were converted to the ion rest frame based on the kinetic energy of the ion beam. This induced an uncertainty in the energy scale that is less than 20~$\mu$eV, which is on the order of the laser bandwidth.

\begin{figure}[tb]
\centering
\includegraphics[width = \figWidth]{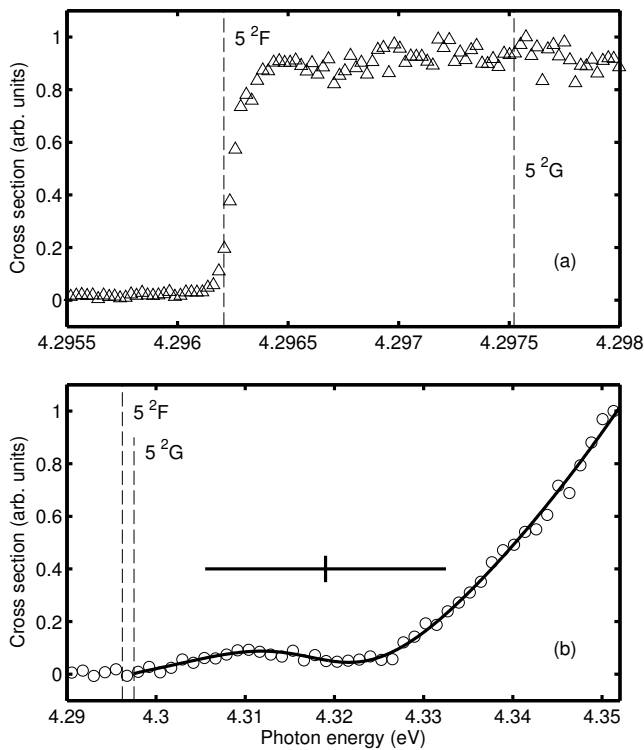}
\caption{
Partial photodetachment cross sections measured in the \Kf\  channel (a) and \Kg\ channel (b). Note that the energy range in (a) is 25 times smaller than in (b). The dashed vertical lines indicate the experimentally known threshold positions \cite{sansonetti:7, andersen:1511}. The solid line in (b) is a fit of the function in  Eq.~(\ref{eq:fitFormula}) with $\sigma_{th}$ from Eq.~(\ref{eq:crossSection}). The resonance position and width, as extracted from the fit, are indicated by the vertical and horizontal bars in (b).
\label{fig:thresholds}}
\end{figure}

Figure \ref{fig:thresholds} shows the thresholds for photodetachment to the \Kf\ and \Kg\ channels in panels (a) and (b), respectively. Note that the energy range in (a) is 25 times smaller than in (b).
Figure \ref{fig:thresholds} (a) covers the fast onset of the cross section in the \Kf\ channel, which plateaus approximately 200 $\mu$eV above the threshold. The small signal just below the channel opening is caused by the finite bandwidth of the laser.
This steplike rise at threshold is the same energy dependence that has been seen in photodetachment experiments leaving the residual atom in a state with high positive polarizability \cite{sandstrom:052707}, as discussed above.
The observed threshold behavior deviates strongly from the Wigner law. Even with the correction term of O'Malley \cite{omalley:1668}, the Wigner law cannot be used to reproduce any part of the data in Fig.~\ref{fig:thresholds} (a).

With the large positive polarizability, the height of the centrifugal barrier in the K$(\fm) + e^-(\ell=2)$ channel is reduced to only 31 $\mu$eV. At detachment energies above this value, the electron can pass above the barrier. Therefore, the cross section is expected to rise very rapidly to its maximum value. This is in good agreement with Fig.~\ref{fig:thresholds} (a), in which the cross section reaches 80 \% of the maximum value at 100 $\mu$eV above threshold.

In sharp contrast, the cross section for the \Kg\ channel, shown in  Fig.~\ref{fig:thresholds} (b), has a very slow onset. Moreover, a broad resonance is seen to modulate the cross section at energies around 4.32~eV. Except for this modulation, the cross section is increasing over the whole observable range. The figure contains data up to 54~meV above threshold, which is just below the K(\p) channel opening.

The presence of the resonance in Fig.~\ref{fig:thresholds} (b) complicates the analysis of the nonresonant threshold behavior. The same resonance appears in the partial cross section for the \Ks\ channel (not shown), which opens at 4.255 eV \cite{andersen:1511, sansonetti:7}.
A fit of a Shore profile to this resonance yielded a resonance energy  E$_\mathrm{r}^\mathrm{\sm} = 4.320(3)$ eV and a width $\Gamma_\mathrm{r}^\mathrm{\sm} = 25(9)$ meV.


The primary interest of this Letter is the nonresonant threshold behavior of the \Kg\ photodetachment channel. Photodetachment into this channel yields an electron emitted primarily as an $f$ wave near threshold and hence the Wigner law predicts a cross section that should scale as  ${E_e}^{7/2}$.
However, due to the large polarizability of the \g\ state, a deviation from the Wigner law on the order of 20 \% is expected at energies as low as $10$ $\mu$eV  above threshold
\cite{ * [{}] [{, Note the error in the sign of the correction factor, see for example ref [9].}] omalley:1668}.

As was the case for the threshold in the \Kf\ channel, the Wigner law with the O'Malley correction fails to reproduce the shape of the onset seen in  Fig.~\ref{fig:thresholds} (b).
 Analysis of the experimental data therefore requires a treatment in which the polarization potential is included in a more sophisticated manner.
As discussed above, modified effective range theories have been developed to describe thresholds in the presence of a large positive dipole polarizability.
To gain a qualitative understanding of the observed cross section we have developed a semiclassical model that takes the negative polarizability into account.

In this model we consider the effective potential $U(r)={\ell(\ell+1)}/{2r^2}-{\alpha}/{2r^4}$ (in atomic units) for an electron moving in the field of an atom with the static dipole polarizability $\alpha$.
Since the polarizability of the \g\ state is negative, this represents a purely repulsive potential.
By comparing the centrifugal and induced dipole terms, one estimates that, with the polarizability calculated by Liu \cite{liu:052715}, the polarization potential dominates for electron energies larger than 1~meV.
Since the energy range in the experiment extends significantly above this value,
we consider the limiting case where the centrifugal potential is neglected.
Furthermore, the repulsive potential spatially separates the classically allowed regions of the initial and final states to such an extent that
it is only the exponential tails of the wave functions, in the classically forbidden region,
that contribute significantly to the transition amplitude.
In the classically forbidden region, the wave function of the free electron in the final state can be expressed by using the semiclassical approximation:
\begin{equation}
\psi_k(r)=\frac{C(E_e)}{r\,\sqrt{p(r)}}\exp\left( -\int_r^{r_0}p(x)\,dx\right),\quad r<r_0,
\label{eq:waveFunk}
\end{equation}
where $k=(2E_e)^{1/2}$ denotes the electron momentum, $p(r)=(|\alpha|/r^4-2E_e)^{1/2}$, $r_0=(|\alpha|/2E_e)^{1/4}$ is the classical turning point, and $C(E_e)\sim
E_e^{-1/4}$ is a normalization coefficient on the $k/2\pi$ momentum scale.
The integral in Eq.~(\ref{eq:waveFunk}) can be expressed in terms of the hypergeometric function $_2F_1\left[-\frac{1}{2},-\frac{1}{4};\frac{3}{4};(\frac{r}{r_0})^4\right]$. Since the region of
significance of the wave function overlap lies at $r<<r_0$, $\psi_k$ can be expanded in a power series of $(r/r_0)^4$.
Taking only the first-order term into account results in an expression for $\psi_k$ with separated $r$ and $E_e$ dependencies.
This form allows calculation of the energy dependence of the photodetachment cross section, which is defined by the integral of the squared transition amplitude over the momentum space of the detached electron.
Since the transition amplitude depends on the energy $E_e$ only through $\psi_k(r)$,
it has the same energy dependence as the exponential tail (\ref{eq:waveFunk}) of the continuum wave function. The integral over the momentum space of final states yields a factor ${E_e}^{1/2}$, which cancels the square of $C(E_e)$.
This results in an analytic function for the cross section of the form
\begin{equation}
\sigma_{\mathrm{th}}\sim\exp\big[D\,{E_e}^{1/4}\big],
\label{eq:crossSection}
\end{equation}
where $D =  2^{11/4}\, \pi^{3/2}\, |\alpha|^{1/4}/\Gamma^2(\frac{1}{4}) \approx 2.850\,|\alpha|^{1/4}$ is a numerical constant. 
Very close to threshold, the centrifugal potential cannot be neglected. The Wigner law should therefore give a better description of the cross section for small energies. 
As mentioned previously, however, the range over which the Wigner law correctly describes the energy dependence of the cross section is very small due to the large polarizability of the \g\ state.
In situations where $\ell = 0$, on the other hand, the induced dipole potential is indeed the longest range interaction and the model is thus expected to be valid at threshold. 
In such cases, Eq.~(\ref{eq:crossSection}) represents the true behavior of the cross section, including the noticeable property that the cross section is finite at threshold.


The data shown in Fig.~\ref{fig:thresholds} (b) have been fitted with a function that
takes the presence of a resonance due to a doubly excited state into account.
The function is a slightly modified expression by Liu and Starace \cite{liu:4647}:
\begin{equation}
\sigma_{\mathrm{pd}} = A\, \sigma_{\mathrm{th}} \left(1+ \frac{\epsilon a +b}{1+ \epsilon^2}\right),
\label{eq:fitFormula}
\end{equation}
where $\epsilon = (E_{h\nu} - E_\mathrm{r})/(\Gamma_\mathrm{r}/2)$. The nonresonant component, $\sigma_{\mathrm{th}}$, is represented by Eq.~(\ref{eq:crossSection}). 
The electron that is emitted in the decay of the doubly excited state experiences the same repulsive polarization potential as electrons emitted directly by photodetachment. 
The influence of the resonance can therefore be reasonably approximated as a product of the threshold law and a Shore profile, as in Eq.~(\ref{eq:fitFormula}).
The  function contains 6 free parameters: the common amplitude factor $A$, the factor $D$ in the exponent, the resonance energy $E_\mathrm{r}$, the resonance width $\Gamma_\mathrm{r}$, and the Shore parameters $a$ and $b$, characterizing the shape of the resonance. 
Tabulated experimental values \cite{sansonetti:7, andersen:1511} were used for the threshold energy in the fitting procedure.
Our model yielded a good fit to the data and the extracted energy and width of the resonance were E$_\mathrm{r}^\mathrm{\gm} = 4.319(4)$~eV and $\Gamma_\mathrm{r}^\mathrm{\gm} = 27(4)$~ meV, respectively. The fit is shown in Fig.~\ref{fig:thresholds} (b) together with a horizontal bar indicating the resonance parameters. The resonance parameters are in agreement with those extracted in the \Ks\ channel and those calculated by Liu \cite{liu:052715}. Table \ref{tab:resonance}  compares the three sets of values. We also performed the fitting using fixed values of
$E_\mathrm{r}$ and $\Gamma_\mathrm{r}$ given by either the fit to the \Ks\ channel data or the  calculation by Liu \cite{liu:052715}. These fits, with only four free parameters, gave curves that are indistinguishable from the fit shown as the solid line in Fig.~\ref{fig:thresholds} (b).

\begin{table}[tb]
\caption{\label{tab:resonance}
A comparison of resonance parameters, with quoted uncertainties representing 1 standard deviation.}
\begin{ruledtabular}
\begin{tabular}{lll}
Origin &
\multicolumn{1}{l}{\textrm{Position, $E_\mathrm{r}$ (eV) }} &
\multicolumn{1}{l}{\textrm{Width, $\Gamma_\mathrm{r}$ (meV)}} \\
\hline
\Kg\ channel & 4.319(4) & 27(4) \\
\Ks\ channel &  4.320(3) & 25(9) \\
Calculation \cite{liu:052715} & 4.32339 & 27.7183\\
\end{tabular}
\end{ruledtabular}
\end{table}

The good fit to the data in Fig.~\ref{fig:thresholds} (b) and the good agreement with resonance parameters in Table \ref{tab:resonance} indicates that the nonresonant background can be well represented by the exponential function in Eq.~(\ref{eq:crossSection}). This function includes the polarizability of the excited atomic state.
This means that it is, in principle, possible to extract an estimate of the polarizability of the \g\ state from the shape of the curve shown in Fig.~\ref{fig:thresholds} (b).
The result from the fit gives a polarizability of $-2.7\times10^4$~a.u., indeed a very large value, although significantly smaller than $-3.1\times10^6$~a.u. as calculated by Liu \cite{liu:052715}.
However, we do not expect our model, which was developed to achieve a qualitative understanding of the threshold behavior, to be able to give  a reliable value of the polarizability.
In particular, to neglect the centrifugal potential and consider only the polarization interaction is most likely an oversimplification. In addition, the simple $1/r^{4} $ radial dependence of the induced dipole potential might not be a complete description of the polarization interaction, especially for small radial distances.
When the interaction between the atom and the free electron is weak, a perturbative treatment results in the simple $1/r^4$ radial dependence. When, on the other hand, the interaction becomes strong, nonperturbative  methods may be needed. This could result in a more complicated radial dependence of the repulsive potential than used in our analysis. It might also be important to take the dynamical effects on the polarizability into account in a full model of the threshold behavior.

In summary, we have investigated near-threshold cross sections for two channels in which the residual atom is left in highly excited and highly polarizable excited states. The channels differ in that the  sign of the dipole polarizability is different.  The induced dipole interaction is so strong that it dominates and determines the general shape of the photodetachment cross sections.
In the case of the \Kf\ channel, the attractive polarization potential reduces the height of the centrifugal barrier. 
This leads to a steep onset of the cross section, which plateaus once the electron can pass above the barrier.
In sharp contrast, the \Kg\ channel exhibits a much slower onset above threshold. This is due to the repulsive potential, which precludes a large overlap of the wave functions at low electron energies.
We have shown that a qualitative understanding of the shape of the cross section can be obtained through a semiclassical model. The resulting exponential energy dependency has been successfully used to represent the nonresonant component in a fit to the data, which also includes a resonance.

It is expected that the data and model presented in this Letter will initiate further theoretical work, resulting in a more detailed understanding of the observed threshold behaviors.
There are many pairs of states in the alkali atoms that have large polarizabilities with opposite signs. It is our intention to systematically probe these states in order to investigate the threshold behaviors as a function of polarizability.

\begin{acknowledgments}
We thank Sten Salomonson for enlightening discussions on polarizability  and induced polarization potentials.
Financial support from the Swedish Research Council is gratefully acknowledged. CWW received support from the Wenner-Gren Foundation, the Andrew W. Mellon Foundation, and NSF Grant No. 0757976.
HH and IK acknowledge the support by the Deutsche Forschungsgemeinshaft, Grant No KI 865/3-1.
\end{acknowledgments}

\bibliography{Artiklar} 

\begin{thebibliography}{22}%
\makeatletter
\providecommand \@ifxundefined [1]{%
 \@ifx{#1\undefined}
}%
\providecommand \@ifnum [1]{%
 \ifnum #1\expandafter \@firstoftwo
 \else \expandafter \@secondoftwo
 \fi
}%
\providecommand \@ifx [1]{%
 \ifx #1\expandafter \@firstoftwo
 \else \expandafter \@secondoftwo
 \fi
}%
\providecommand \natexlab [1]{#1}%
\providecommand \enquote  [1]{``#1''}%
\providecommand \bibnamefont  [1]{#1}%
\providecommand \bibfnamefont [1]{#1}%
\providecommand \citenamefont [1]{#1}%
\providecommand \href@noop [0]{\@secondoftwo}%
\providecommand \href [0]{\begingroup \@sanitize@url \@href}%
\providecommand \@href[1]{\@@startlink{#1}\@@href}%
\providecommand \@@href[1]{\endgroup#1\@@endlink}%
\providecommand \@sanitize@url [0]{\catcode `\\12\catcode `\$12\catcode
  `\&12\catcode `\#12\catcode `\^12\catcode `\_12\catcode `\%12\relax}%
\providecommand \@@startlink[1]{}%
\providecommand \@@endlink[0]{}%
\providecommand \url  [0]{\begingroup\@sanitize@url \@url }%
\providecommand \@url [1]{\endgroup\@href {#1}{\urlprefix }}%
\providecommand \urlprefix  [0]{URL }%
\providecommand \Eprint [0]{\href }%
\providecommand \doibase [0]{http://dx.doi.org/}%
\providecommand \selectlanguage [0]{\@gobble}%
\providecommand \bibinfo  [0]{\@secondoftwo}%
\providecommand \bibfield  [0]{\@secondoftwo}%
\providecommand \translation [1]{[#1]}%
\providecommand \BibitemOpen [0]{}%
\providecommand \bibitemStop [0]{}%
\providecommand \bibitemNoStop [0]{.\EOS\space}%
\providecommand \EOS [0]{\spacefactor3000\relax}%
\providecommand \BibitemShut  [1]{\csname bibitem#1\endcsname}%
\let\auto@bib@innerbib\@empty
\bibitem [{\citenamefont {Einstein}(1905)}]{einstein:132}%
  \BibitemOpen
  \bibfield  {author} {\bibinfo {author} {\bibfnamefont {A.}~\bibnamefont
  {Einstein}},\ }\href {\doibase 10.1002/andp.19053220607} {\bibfield
  {journal} {\bibinfo  {journal} {Annalen der Physik}\ }\textbf {\bibinfo
  {volume} {322}},\ \bibinfo {pages} {132} (\bibinfo {year}
  {1905})}\BibitemShut {NoStop}%
\bibitem [{\citenamefont {Sadeghpour}\ \emph {et~al.}(2000)\citenamefont
  {Sadeghpour}, \citenamefont {Bohn}, \citenamefont {Cavagnero}, \citenamefont
  {Esry}, \citenamefont {Fabrikant}, \citenamefont {Macek},\ and\ \citenamefont
  {Rau}}]{sadeghpour:R93}%
  \BibitemOpen
  \bibfield  {author} {\bibinfo {author} {\bibfnamefont {H.~R.}\ \bibnamefont
  {Sadeghpour}}, \bibinfo {author} {\bibfnamefont {J.~L.}\ \bibnamefont
  {Bohn}}, \bibinfo {author} {\bibfnamefont {M.~J.}\ \bibnamefont {Cavagnero}},
  \bibinfo {author} {\bibfnamefont {B.~D.}\ \bibnamefont {Esry}}, \bibinfo
  {author} {\bibfnamefont {I.~I.}\ \bibnamefont {Fabrikant}}, \bibinfo {author}
  {\bibfnamefont {J.~H.}\ \bibnamefont {Macek}}, \ and\ \bibinfo {author}
  {\bibfnamefont {A.~R.~P.}\ \bibnamefont {Rau}},\ }\href
  {http://stacks.iop.org/0953-4075/33/i=5/a=201} {\bibfield  {journal}
  {\bibinfo  {journal} {J. Phys. B: At. Mol. Opt. Phys.}\ }\textbf {\bibinfo
  {volume} {33}},\ \bibinfo {pages} {R93} (\bibinfo {year} {2000})}\BibitemShut
  {NoStop}%
\bibitem [{\citenamefont {Jonson}(2004)}]{jonson:1}%
  \BibitemOpen
  \bibfield  {author} {\bibinfo {author} {\bibfnamefont {B.}~\bibnamefont
  {Jonson}},\ }\href {\doibase DOI: 10.1016/j.physrep.2003.07.004} {\bibfield
  {journal} {\bibinfo  {journal} {Phys. Rep.}\ }\textbf {\bibinfo {volume}
  {389}},\ \bibinfo {pages} {1 } (\bibinfo {year} {2004})}\BibitemShut
  {NoStop}%
\bibitem [{\citenamefont {DeMarco}\ \emph {et~al.}(1999)\citenamefont
  {DeMarco}, \citenamefont {Bohn}, \citenamefont {Burke}, \citenamefont
  {Holland},\ and\ \citenamefont {Jin}}]{demarco:4208}%
  \BibitemOpen
  \bibfield  {author} {\bibinfo {author} {\bibfnamefont {B.}~\bibnamefont
  {DeMarco}}, \bibinfo {author} {\bibfnamefont {J.~L.}\ \bibnamefont {Bohn}},
  \bibinfo {author} {\bibfnamefont {J.~P.}\ \bibnamefont {Burke}}, \bibinfo
  {author} {\bibfnamefont {M.}~\bibnamefont {Holland}}, \ and\ \bibinfo
  {author} {\bibfnamefont {D.~S.}\ \bibnamefont {Jin}},\ }\href {\doibase
  10.1103/PhysRevLett.82.4208} {\bibfield  {journal} {\bibinfo  {journal}
  {Phys. Rev. Lett.}\ }\textbf {\bibinfo {volume} {82}},\ \bibinfo {pages}
  {4208} (\bibinfo {year} {1999})}\BibitemShut {NoStop}%
\bibitem [{\citenamefont {Zhang}\ \emph {et~al.}(2004)\citenamefont {Zhang},
  \citenamefont {Cheng}, \citenamefont {Kim}, \citenamefont {Stanojevic},\ and\
  \citenamefont {Eyler}}]{zhang:203003}%
  \BibitemOpen
  \bibfield  {author} {\bibinfo {author} {\bibfnamefont {Y.~P.}\ \bibnamefont
  {Zhang}}, \bibinfo {author} {\bibfnamefont {C.~H.}\ \bibnamefont {Cheng}},
  \bibinfo {author} {\bibfnamefont {J.~T.}\ \bibnamefont {Kim}}, \bibinfo
  {author} {\bibfnamefont {J.}~\bibnamefont {Stanojevic}}, \ and\ \bibinfo
  {author} {\bibfnamefont {E.~E.}\ \bibnamefont {Eyler}},\ }\href {\doibase
  10.1103/PhysRevLett.92.203003} {\bibfield  {journal} {\bibinfo  {journal}
  {Phys. Rev. Lett.}\ }\textbf {\bibinfo {volume} {92}},\ \bibinfo {pages}
  {203003} (\bibinfo {year} {2004})}\BibitemShut {NoStop}%
\bibitem [{\citenamefont {Rupak}\ and\ \citenamefont
  {Higa}(2011)}]{rupak:222501}%
  \BibitemOpen
  \bibfield  {author} {\bibinfo {author} {\bibfnamefont {G.}~\bibnamefont
  {Rupak}}\ and\ \bibinfo {author} {\bibfnamefont {R.}~\bibnamefont {Higa}},\
  }\href {\doibase 10.1103/PhysRevLett.106.222501} {\bibfield  {journal}
  {\bibinfo  {journal} {Phys. Rev. Lett.}\ }\textbf {\bibinfo {volume} {106}},\
  \bibinfo {pages} {222501} (\bibinfo {year} {2011})}\BibitemShut {NoStop}%
\bibitem [{\citenamefont {Wigner}(1948)}]{wigner:1002}%
  \BibitemOpen
  \bibfield  {author} {\bibinfo {author} {\bibfnamefont {E.~P.}\ \bibnamefont
  {Wigner}},\ }\href {\doibase 10.1103/PhysRev.73.1002} {\bibfield  {journal}
  {\bibinfo  {journal} {Phys. Rev.}\ }\textbf {\bibinfo {volume} {73}},\
  \bibinfo {pages} {1002} (\bibinfo {year} {1948})}\BibitemShut {NoStop}%
\bibitem [{\citenamefont {Bilodeau}\ \emph {et~al.}(1999)\citenamefont
  {Bilodeau}, \citenamefont {Scheer}, \citenamefont {Haugen},\ and\
  \citenamefont {Brooks}}]{bilodeau:012505}%
  \BibitemOpen
  \bibfield  {author} {\bibinfo {author} {\bibfnamefont {R.~C.}\ \bibnamefont
  {Bilodeau}}, \bibinfo {author} {\bibfnamefont {M.}~\bibnamefont {Scheer}},
  \bibinfo {author} {\bibfnamefont {H.~K.}\ \bibnamefont {Haugen}}, \ and\
  \bibinfo {author} {\bibfnamefont {R.~L.}\ \bibnamefont {Brooks}},\ }\href
  {\doibase 10.1103/PhysRevA.61.012505} {\bibfield  {journal} {\bibinfo
  {journal} {Phys. Rev. A}\ }\textbf {\bibinfo {volume} {61}},\ \bibinfo
  {pages} {012505} (\bibinfo {year} {1999})}\BibitemShut {NoStop}%
\bibitem [{\citenamefont {Hotop}\ \emph {et~al.}(1973)\citenamefont {Hotop},
  \citenamefont {Patterson},\ and\ \citenamefont {Lineberger}}]{hotop:762}%
  \BibitemOpen
  \bibfield  {author} {\bibinfo {author} {\bibfnamefont {H.}~\bibnamefont
  {Hotop}}, \bibinfo {author} {\bibfnamefont {T.~A.}\ \bibnamefont
  {Patterson}}, \ and\ \bibinfo {author} {\bibfnamefont {W.~C.}\ \bibnamefont
  {Lineberger}},\ }\href {\doibase 10.1103/PhysRevA.8.762} {\bibfield
  {journal} {\bibinfo  {journal} {Phys. Rev. A}\ }\textbf {\bibinfo {volume}
  {8}},\ \bibinfo {pages} {762} (\bibinfo {year} {1973})}\BibitemShut {NoStop}%
\bibitem [{\citenamefont {Hotop}\ and\ \citenamefont
  {Lineberger}(1973)}]{hotop:2379}%
  \BibitemOpen
  \bibfield  {author} {\bibinfo {author} {\bibfnamefont {H.}~\bibnamefont
  {Hotop}}\ and\ \bibinfo {author} {\bibfnamefont {W.~C.}\ \bibnamefont
  {Lineberger}},\ }\href {\doibase 10.1063/1.1679515} {\bibfield  {journal}
  {\bibinfo  {journal} {J Chem. Phys.}\ }\textbf {\bibinfo {volume} {58}},\
  \bibinfo {pages} {2379} (\bibinfo {year} {1973})}\BibitemShut {NoStop}%
\bibitem [{\citenamefont {Bilodeau}\ \emph {et~al.}(2005)\citenamefont
  {Bilodeau}, \citenamefont {Bozek}, \citenamefont {Gibson}, \citenamefont
  {Walter}, \citenamefont {Ackerman}, \citenamefont {Dumitriu},\ and\
  \citenamefont {Berrah}}]{bilodeau:083001}%
  \BibitemOpen
  \bibfield  {author} {\bibinfo {author} {\bibfnamefont {R.~C.}\ \bibnamefont
  {Bilodeau}}, \bibinfo {author} {\bibfnamefont {J.~D.}\ \bibnamefont {Bozek}},
  \bibinfo {author} {\bibfnamefont {N.~D.}\ \bibnamefont {Gibson}}, \bibinfo
  {author} {\bibfnamefont {C.~W.}\ \bibnamefont {Walter}}, \bibinfo {author}
  {\bibfnamefont {G.~D.}\ \bibnamefont {Ackerman}}, \bibinfo {author}
  {\bibfnamefont {I.}~\bibnamefont {Dumitriu}}, \ and\ \bibinfo {author}
  {\bibfnamefont {N.}~\bibnamefont {Berrah}},\ }\href {\doibase
  10.1103/PhysRevLett.95.083001} {\bibfield  {journal} {\bibinfo  {journal}
  {Phys. Rev. Lett.}\ }\textbf {\bibinfo {volume} {95}},\ \bibinfo {pages}
  {083001} (\bibinfo {year} {2005})}\BibitemShut {NoStop}%
\bibitem [{\citenamefont {Farley}(1989)}]{farley:6286}%
  \BibitemOpen
  \bibfield  {author} {\bibinfo {author} {\bibfnamefont {J.~W.}\ \bibnamefont
  {Farley}},\ }\href {\doibase 10.1103/PhysRevA.40.6286} {\bibfield  {journal}
  {\bibinfo  {journal} {Phys. Rev. A}\ }\textbf {\bibinfo {volume} {40}},\
  \bibinfo {pages} {6286} (\bibinfo {year} {1989})}\BibitemShut {NoStop}%
\bibitem [{\citenamefont {O'Malley}(1965)}]{omalley:1668}%
  \BibitemOpen
  \bibfield  {author} {\bibinfo {author} {\bibfnamefont {T.~F.}\ \bibnamefont
  {O'Malley}},\ }\href {\doibase 10.1103/PhysRev.137.A1668} {\bibfield
  {journal} {\bibinfo  {journal} {Phys. Rev.}\ }\textbf {\bibinfo {volume}
  {137}},\ \bibinfo {pages} {A1668} (\bibinfo {year} {1965})}\BibitemShut
  {NoStop}%
\bibitem [{\citenamefont {Watanabe}\ and\ \citenamefont
  {Greene}(1980)}]{watanabe:158}%
  \BibitemOpen
  \bibfield  {author} {\bibinfo {author} {\bibfnamefont {S.}~\bibnamefont
  {Watanabe}}\ and\ \bibinfo {author} {\bibfnamefont {C.~H.}\ \bibnamefont
  {Greene}},\ }\href {\doibase 10.1103/PhysRevA.22.158} {\bibfield  {journal}
  {\bibinfo  {journal} {Phys. Rev. A}\ }\textbf {\bibinfo {volume} {22}},\
  \bibinfo {pages} {158} (\bibinfo {year} {1980})}\BibitemShut {NoStop}%
\bibitem [{\citenamefont {Sandstr\"om}\ \emph {et~al.}(2004)\citenamefont
  {Sandstr\"om}, \citenamefont {Haeffler}, \citenamefont {Kiyan}, \citenamefont
  {Berzinsh}, \citenamefont {Hanstorp}, \citenamefont {Pegg}, \citenamefont
  {Hunnell},\ and\ \citenamefont {Ward}}]{sandstrom:052707}%
  \BibitemOpen
  \bibfield  {author} {\bibinfo {author} {\bibfnamefont {J.}~\bibnamefont
  {Sandstr\"om}}, \bibinfo {author} {\bibfnamefont {G.}~\bibnamefont
  {Haeffler}}, \bibinfo {author} {\bibfnamefont {I.}~\bibnamefont {Kiyan}},
  \bibinfo {author} {\bibfnamefont {U.}~\bibnamefont {Berzinsh}}, \bibinfo
  {author} {\bibfnamefont {D.}~\bibnamefont {Hanstorp}}, \bibinfo {author}
  {\bibfnamefont {D.~J.}\ \bibnamefont {Pegg}}, \bibinfo {author}
  {\bibfnamefont {J.~C.}\ \bibnamefont {Hunnell}}, \ and\ \bibinfo {author}
  {\bibfnamefont {S.~J.}\ \bibnamefont {Ward}},\ }\href {\doibase
  10.1103/PhysRevA.70.052707} {\bibfield  {journal} {\bibinfo  {journal} {Phys.
  Rev. A}\ }\textbf {\bibinfo {volume} {70}},\ \bibinfo {pages} {052707}
  (\bibinfo {year} {2004})}\BibitemShut {NoStop}%
\bibitem [{\citenamefont {Liu}(2001)}]{liu:052715}%
  \BibitemOpen
  \bibfield  {author} {\bibinfo {author} {\bibfnamefont {C.-N.}\ \bibnamefont
  {Liu}},\ }\href {\doibase 10.1103/PhysRevA.64.052715} {\bibfield  {journal}
  {\bibinfo  {journal} {Phys. Rev. A}\ }\textbf {\bibinfo {volume} {64}},\
  \bibinfo {pages} {052715} (\bibinfo {year} {2001})}\BibitemShut {NoStop}%
\bibitem [{\citenamefont {Sansonetti}(2008)}]{sansonetti:7}%
  \BibitemOpen
  \bibfield  {author} {\bibinfo {author} {\bibfnamefont {J.~E.}\ \bibnamefont
  {Sansonetti}},\ }\href {\doibase 10.1063/1.2789451} {\bibfield  {journal}
  {\bibinfo  {journal} {J. Phys. Chem. Ref. Data}\ }\textbf {\bibinfo {volume}
  {37}},\ \bibinfo {pages} {7} (\bibinfo {year} {2008})}\BibitemShut {NoStop}%
\bibitem [{\citenamefont {Blatt}\ and\ \citenamefont
  {Weisskopf}(1979)}]{blatt:weisskopf}%
  \BibitemOpen
  \bibfield  {author} {\bibinfo {author} {\bibfnamefont {J.~M.}\ \bibnamefont
  {Blatt}}\ and\ \bibinfo {author} {\bibfnamefont {V.~F.}\ \bibnamefont
  {Weisskopf}},\ }\href@noop {} {\emph {\bibinfo {title} {Theoretical Nuclear
  Physics}}}\ (\bibinfo  {publisher} {Springer-Verlag},\ \bibinfo {address}
  {New York},\ \bibinfo {year} {1979})\BibitemShut {NoStop}%
\bibitem [{\citenamefont {Diehl}\ \emph {et~al.}(2011)\citenamefont {Diehl},
  \citenamefont {Wendt}, \citenamefont {Lindahl}, \citenamefont {Andersson},\
  and\ \citenamefont {Hanstorp}}]{diehl:053302}%
  \BibitemOpen
  \bibfield  {author} {\bibinfo {author} {\bibfnamefont {C.}~\bibnamefont
  {Diehl}}, \bibinfo {author} {\bibfnamefont {K.}~\bibnamefont {Wendt}},
  \bibinfo {author} {\bibfnamefont {A.~O.}\ \bibnamefont {Lindahl}}, \bibinfo
  {author} {\bibfnamefont {P.}~\bibnamefont {Andersson}}, \ and\ \bibinfo
  {author} {\bibfnamefont {D.}~\bibnamefont {Hanstorp}},\ }\href {\doibase
  10.1063/1.3587617} {\bibfield  {journal} {\bibinfo  {journal} {Rev. Sci.
  Instrum.}\ }\textbf {\bibinfo {volume} {82}},\ \bibinfo {eid} {053302}
  (\bibinfo {year} {2011})}\BibitemShut {NoStop}%
\bibitem [{\citenamefont {Dahl}(2000)}]{dahl:3}%
  \BibitemOpen
  \bibfield  {author} {\bibinfo {author} {\bibfnamefont {D.~A.}\ \bibnamefont
  {Dahl}},\ }\href {\doibase DOI: 10.1016/S1387-3806(00)00305-5} {\bibfield
  {journal} {\bibinfo  {journal} {Int. J. Mass Spectrom.}\ }\textbf {\bibinfo
  {volume} {200}},\ \bibinfo {pages} {3 } (\bibinfo {year} {2000})}\BibitemShut
  {NoStop}%
\bibitem [{\citenamefont {Andersen}\ \emph {et~al.}(1999)\citenamefont
  {Andersen}, \citenamefont {Haugen},\ and\ \citenamefont
  {Hotop}}]{andersen:1511}%
  \BibitemOpen
  \bibfield  {author} {\bibinfo {author} {\bibfnamefont {T.}~\bibnamefont
  {Andersen}}, \bibinfo {author} {\bibfnamefont {H.~K.}\ \bibnamefont
  {Haugen}}, \ and\ \bibinfo {author} {\bibfnamefont {H.}~\bibnamefont
  {Hotop}},\ }\href {\doibase 10.1063/1.556047} {\bibfield  {journal} {\bibinfo
   {journal} {J. Phys. Chem. Ref. Data}\ }\textbf {\bibinfo {volume} {28}},\
  \bibinfo {pages} {1511} (\bibinfo {year} {1999})}\BibitemShut {NoStop}%
\bibitem [{\citenamefont {Liu}\ and\ \citenamefont {Starace}(1999)}]{liu:4647}%
  \BibitemOpen
  \bibfield  {author} {\bibinfo {author} {\bibfnamefont {C.-N.}\ \bibnamefont
  {Liu}}\ and\ \bibinfo {author} {\bibfnamefont {A.~F.}\ \bibnamefont
  {Starace}},\ }\href {\doibase 10.1103/PhysRevA.60.4647} {\bibfield  {journal}
  {\bibinfo  {journal} {Phys. Rev. A}\ }\textbf {\bibinfo {volume} {60}},\
  \bibinfo {pages} {4647} (\bibinfo {year} {1999})}\BibitemShut {NoStop}%
\end{thebibliography}%

\end{document}